\documentclass[fleqn,10pt]{wlscirep}
\usepackage[utf8]{inputenc}
\usepackage[T1]{fontenc}
\usepackage{xcolor}

\usepackage[utf8]{inputenc} 
\usepackage[T1]{fontenc}    
\usepackage{hyperref}       
\usepackage{url}            
\usepackage{booktabs}       
\usepackage{amsfonts}       
\usepackage{nicefrac}       
\usepackage{microtype}      
\usepackage{graphicx}
\usepackage{doi}
\usepackage{subcaption}
\usepackage{makecell}
\usepackage{chngpage}
\usepackage{gensymb}
\usepackage{tablefootnote}
\usepackage{xurl}
\usepackage{mathtools}
\usepackage{xcolor}

\title{Data fusion of complementary data sources using Machine Learning enables higher accuracy Solar Resource Maps}

\author[1,*]{J. Rabault}  
\author[1,2]{M. L. Sætra} 
\author[1]{A. Dobler}     
\author[1]{S. Eastwood}
\author[1]{E. Berge}   

\affil[1]{Norwegian Meteorological Institute, Oslo, Norway}
\affil[2]{Department of Computer Science, Oslo Metropolitan University, Oslo, Norway}

\affil[*]{corresponding author: Jean Rabault (jean.rblt@gmail.com)}

\begin{abstract}
In the present work, we collect solar irradiance and atmospheric condition data from several products, obtained from both numerical models (ERA5 and NORA3) and satellite observations (CMSAF-SARAH3). We then train simple supervised Machine Learning (ML) data fusion models, using these products as predictors and direct in-situ Global Horizontal Irradiance (GHI) measurements over Norway as ground-truth. We show that combining these products by applying our trained ML models provides a GHI estimate that is significantly more accurate than that obtained from any product taken individually. Using the trained models, we generate a 30-year ML-corrected map of GHI over Norway, which we release as a new open data product. Our ML-based data fusion methodology could be applied, after suitable training and input data selection, to any geographic area on Earth.
\end{abstract}

\begin{document}
\flushbottom
\maketitle

\section{Introduction}

\begin{figure*}
    \centering
    \includegraphics[width=0.95\linewidth]{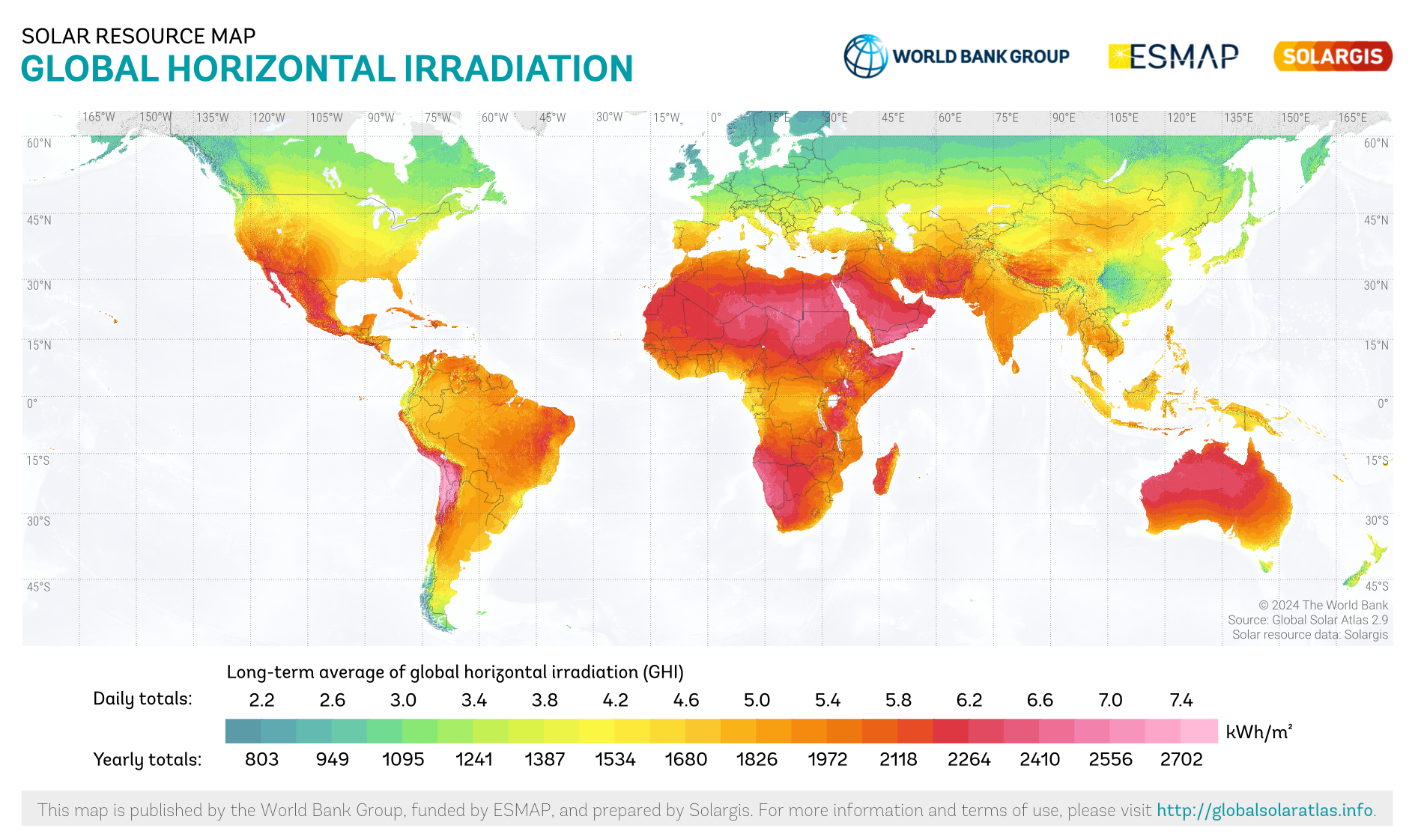}
    \caption{Exported Global Solar Atlas map (\url{https://globalsolaratlas.info}, recovered 2025-01-09). This provides a global overview of the average daily and yearly-accumulated Global Horizontal Irradiance (GHI), in kWh/m$^2$. Note that data are typically available only within the latitude range [65S; 65N], owing to the use of geostationary satellite data to generate the product.}
    \label{fig:GlobalSolarAtlas}
\end{figure*}

The mapping of solar energy resources is attracting increased attention following the evolution in electricity production systems deployed throughout the world \cite{gueymard2012clear,polo2015solar,wegertseder2016combining,pruavualie2019spatial,brent2020solar}. Global products are available and give a general overview of global resource availability and distribution, for example, the Global Solar Atlas (GSA) \cite{suri2019global}, which is available at \url{https://globalsolaratlas.info}, see the illustration in Fig. \ref{fig:GlobalSolarAtlas}. However, these global products must be complemented by high accuracy, fine resolution, locally optimized resource maps over specific regions to support decisions about the deployment of electrical production systems in a given area \cite{castillo2016assessment}. Moreover, global products such as the GSA may have limitations in specific regions. For example, the GSA uses data from geostationary satellites to generate its estimate \cite{perez2013semi}, and therefore does not provide information poleward of approximately 65\degree{}, making it impossible to use over, e.g., large areas of northern Norway. 

Therefore, the development of optimized regional solar resource maps is a topic of interest. The interest of such maps is double: i) they can cover areas that may not be as well covered by global products, as highlighted above, and ii) they can be tuned and optimized to provide better results locally, by taking into account local climatology and local model biases that depend on both the models that are run over the area and the dominating physics, weather and climate patterns. {\color{black}In the realm of optimized solar potential quantification, site adaptation is a common technique that provides higher accuracy solar resource estimates at a single location in space based on short-term in-situ time series measurements. This can rely on either classical statistical methods, or machine learning methods, or a combination of both, see e.g. \cite{cebecauer2016site,fernandez2020site,yang2021probabilistic,narvaez2021machine,ruiz2024accurate,zainali2024site}. In the present work, by contrast, we use a set of spatially distributed solar radiation measurement stations that have gathered data over a period of several years to train Machine Learning (ML) models that improve the solar resource estimate over a whole geographic area. As a consequence, our work presents some similarities with site adaptation, since both methods aim at generating higher quality solar radiation estimates, but the scope and methodology we employ here is slightly different from what has been developed by the site adaptation community. These two methodologies can complement each others: for example, one could use our present methodology to generate an optimized solar resource map over a geographic region of interest, and use it as the initial model timeseries input for performing site adaptation.}

More specifically, we describe a methodology for generating Global Horizontal Irradiance (GHI) estimates over Norway by using simple ML models to combine data from several input products. We analyze the performance of our models, and we then describe how these can be used to generate a 30-year (1991--2020) best estimate GHI map over Norway at 0.03\degree{} (approximately 3.3 km) spatial- and 1-hour time-resolution, and we provide access to this product on the THREDDS Data Server (TDS) of the Norwegian Meteorological Institute (MET Norway). This paper is an extended discussion of what is succinctly described in the end-user manual for the GHI map, which was previously published at \url{https://github.com/metno/2024_Sunpoint_solar_irradiance_map_over_Norway_1991-2020}. 

The paper is organized as follows. We first describe the data sources used to train, validate, and deploy our ML models. Then, we describe our ML methodology and the different models that we train. Finally, we present our results and compare the accuracy of the model predictions with the different data sources available over Norway. We conclude by discussing the main findings and impact of this study and providing a perspective on how these data can be used to map the potential for solar energy generation in Norway, and how our present methodology could be further improved and applied to other areas.

\section{Data sources}

In this section, we discuss the different data sources and products used in the present study. We use two main categories of data in this work: i) products providing estimates of incoming solar energy conditions, atmosphere conditions, and ground conditions over Norway that are used as input to the ML models, and ii) direct ground-truth in-situ measurements of the GHI obtained from pyranometers deployed in Norway, which are used as a target to train and validate ML models. Within the products used in i), there are two subcategories: ia) model products that are obtained from numerical weather models, and ib) satellite-based products. In the following, we provide a detailed overview of each of these data categories.

The main requirements for choosing data sources for category i) to be used in the present study are as follows:

\begin{itemize}
    \item R1) we need that the data are produced over the full 30-year period (1991--2020), so that we can generate our final product, i.e., a solar resource map over Norway covering this full 30-year period,
    \item R2) the data generated need to be as consistent as possible over this 30-year time period, and in particular this implies that there should be no major changes in the properties of the data used over this 30-years period.
\end{itemize}

Requirement R1) is a direct consequence of our goal of generating a new product that covers the 30-year period. Requirement R2) is due to the methodology used here. Since we use ML models taking a variety of data sources as input to produce our GHI map, we need to make sure that the corresponding data sources have properties that are as constant as possible over the full time period. Indeed, we expect that the ML model helps generate higher-quality GHI estimates by precisely identifying and compensating for biases and imperfections that are systematically present in the input products. Therefore, properties such as model imperfections and biases should be as constant as possible over the target 30-year time period.

Unfortunately, the second requirement forces us to discard many operational model datasets. Indeed, many operational models, such as the AROME-MetCoOp weather model \cite{muller2017arome} (which would otherwise be a strong candidate to be used as an input product), are undergoing continuous update and improvement, which makes their errors and biases non-consistent in time. Moreover, for the specific case of AROME-MetCoOp, this model is currently not rerun after each update cycle in hindcast or reanalysis mode over the full target period of the present study, due to labor, computational, and storage costs.

\subsection{Numerical model data}

Numerical model data are a convenient input product in case they are available for the full 30-year period over which we generate the resource map, and there are relevant products that cover the whole spatial domain of interest. Unfortunately, requirement R2) implies that many high-quality operational models that are not rerun in hindcast or reanalysis mode upon model update cannot be used here since, as discussed above, this leads to changes in the properties (including possible biases and error distribution) of the models.

Therefore, in this study, we use only two numerical model products as input for the ML models:

\begin{itemize}
    \item ERA5 \cite{hersbach2020era5}: this is the fifth-generation ECMWF reanalysis for global climate and weather, which is available globally over the period 1940--present. ERA5 data are available at approximately 31 km spatial resolution and 1-hour temporal resolution.
    \item NORA3 \cite{haakenstad2021nora3}: this is a high-resolution numerical mesoscale weather reanalysis dataset from MET Norway. NORA3 data are available at 3 km spatial resolution and 1-hour temporal resolution.
\end{itemize}

We use several variables from both ERA5 and NORA3, as summarized in Table \ref{tab:model_predictors}. As presented there, we collect variables that contain information about i) the incoming solar energy both at the top of the atmosphere and at the surface, ii) the atmospheric conditions focusing on cloud properties (cloudiness, height of the clouds, cloud types i.e. convective or stratiform), and iii) the ground conditions, i.e. surface albedo, type of vegetation and precipitation type. Data fields belonging to the category i) are naturally useful for the present task, as they constitute the models best prediction for either the GHI, or incoming solar radiation at the top of the atmosphere. Data fields ii) contain information about atmospheric conditions that influence the transfer of solar radiation such as absorption, scattering  and reflection. Finally, data fields iii) contain information about the surface conditions that influences how much solar energy is reflected by the ground into the atmosphere, and hence contributes to indirect reflected irradiance (and to errors and uncertainties in the satellite products). Moreover, this surface state information is also potentially useful in combination with the satellite data presented in the next section, since the accuracy and bias of these satellite data is expected to be dependent on the state of snow and ice coverage.

Although numerical models are taking great care to estimate quantities such as GHI as accurately as possible, it is challenging to accurately model all the physics involved in determining the GHI, so we can expect that there are still biases, uncertainties, and inaccuracies in the GHI estimates the models provide, which will be shown in the results later in this paper, and as previously reported by Perez et al.~\cite{perez2013semi}.

\begin{table*}[h!]
\begin{center}
\begin{tabular}{| c | c | c |}
	\hline
	Source model & NetCDF variable name & unit \\
	\hline
	\hline
	 & integral\_of\_surface\_net\_downward\_shortwave\_flux\_wrt\_time & {\color{black}$Ws/m^2$} \\
	 & integral\_of\_toa\_net\_downward\_shortwave\_flux\_wrt\_time & {\color{black}$Ws/m^2$} \\
	 & air\_temperature\_2m & {\color{black}$K$} \\
	 & low\_type\_cloud\_area\_fraction & {\color{black}1} \\
	 & cloud\_area\_fraction & {\color{black}1} \\
	 & lwe\_thickness\_of\_atmosphere\_mass\_content\_of\_water\_vapor & {\color{black}$m$} \\
	 & convective\_cloud\_area\_fraction & {\color{black}1} \\
	NORA3 & medium\_type\_cloud\_area\_fraction & {\color{black}1} \\
	 & high\_type\_cloud\_area\_fraction & {\color{black}1} \\
	 & precipitation\_amount\_acc & {\color{black}$kg/m^2$} \\
	 & integral\_of\_surface\_downwelling\_shortwave\_flux\_in\_air\_wrt\_time & {\color{black}$Ws/m^2$} \\
	 & snowfall\_amount\_acc & {\color{black}$kg/m^2$} \\
	 & snow\_albedo & {\color{black}1} \\
	 & surface\_total\_albedo & {\color{black}1} \\
	 & leaf\_area\_index & {\color{black}1} \\
	 & vegetation\_fraction & {\color{black}1} \\
    \hline
	 & top\_net\_solar\_radiation & {\color{black} $J/m^2$} \\
	 & surface\_net\_solar\_radiation & {\color{black}$J/m^2$} \\
	 & surface\_solar\_radiation\_downwards & {\color{black}$J/m^2$} \\
	 & total\_cloud\_cover & {\color{black}1} \\
	 & high\_cloud\_cover & {\color{black}1} \\
	 & medium\_cloud\_cover & {\color{black}1} \\
	ERA5 & low\_cloud\_cover & {\color{black}1} \\
	 & snowfall & {\color{black}$m$ (eq. water)} \\
	 & total\_precipitation & {\color{black}$m$} \\
	 & snow\_albedo & {\color{black}1} \\
	 & forecast\_albedo & {\color{black}1} \\
	 & leaf\_area\_index\_low\_vegetation & {\color{black}$m^2 / m^2$} \\
	 & leaf\_area\_index\_high\_vegetation & {\color{black}$m^2 / m^2$} \\
	 & total\_column\_water\_vapour & {\color{black}$kg / m^2$} \\

	\hline
 	\hline
\end{tabular}
    \caption{The list of model variables used as input to the GHI models trained in the following. The variables are collected from both the NORA3 and ERA5 models. The list of variables used here was determined based on discussions with domain experts, who pinpointed what predictors can be relevant to develop a GHI model, as discussed in the text. The variables are fully documented in the ERA5 and the NORA3 netCDF data files \cite{eaton2003netcdf}, see \url{https://cds.climate.copernicus.eu/datasets/reanalysis-era5-single-levels} and \url{https://cfconventions.org/Data/cf-standard-names/current/build/cf-standard-name-table.html}, for further details. {\color{black}The unit choices reflect the content of the source data files, and do not play a role in the actual ML models developed below, since these models can easily normalize, renormalize, and denormalize inputs and outputs.}}
    \label{tab:model_predictors}
\end{center}
\end{table*}

We do not include information on the latitude, longitude, or altitude of the stations, nor on the time or date. This is because the number of pyranometers in Norway, and hence the number of locations used for training, is limited in both time and space. Therefore, including this information (which would otherwise be relevant) may lead to ML algorithms partially managing to identify each station individually and, hence, overfitting. The influence of these parameters on relevant quantities is still available to the ML models through their influence on the predictors, so that we expect that not providing these parameters has only a weak influence on the performance of the models.

The model data are nearest-neighbor interpolated to the position of the pyranometers to generate the training dataset. When generating the final product, these data are nearest-neighbor interpolated to the grid points where each GHI prediction is performed.

\subsection{Satellite data}

Satellites are key tools for Earth monitoring \cite{aschbacher2012european}, including measuring solar irradiance \cite{noia1993solar,huang2019estimating}. We use solar irradiance measurements from geostationary satellites as input data products. The advantage of these data is that they are available regularly (every 30 minutes for the product selected here) in the entire satellite coverage area. This makes geostationary data easy to incorporate into ML models that need collocated input data in space and time to perform predictions. The downside is that geostationary satellites are not well suited to measure irradiance further north than 65\degree{} north (for the product used over Norway) because of the high satellite viewing angle at these high latitudes. This means that satellite data are not available for the northern half of Norway. We do not consider the use of polar orbiting satellites in this study, due to the challenges associated with collocating their swaths in time and space with other measurements and the irregular time coverage at any given location.

More specifically, we use geostationary measurements of solar irradiance as provided by the EUMETSAT CM SAF SARAH-3 product \cite{https://doi.org/10.5676/eum_saf_cm/sarah/v003}. The data are openly available from \url{https://navigator.eumetsat.int/product/EO:EUM:DAT:0863}, at a time resolution of 30 minutes and a spatial resolution of 0.05\degree{}. Using the SARAH-3 dataset ensures that, while the actual satellites being used and the instruments they carry vary within the 30-year period, the data processing methodology is homogeneous over the full time span considered. In particular, best efforts are made by the data producer to avoid sensor drift in time (see the self calibration procedure highlighted in \cite{algo_cmsaf, posselt2012remote}). When either training or prediction are performed at a given hour $H$, we provide as input to the models the three closest SARAH-3 estimates, corresponding to the set of time instants $\{H$ - 30 min, $H$, $H$ + 30 min$\}$. Interpolation to the location of the pyranometers (which is performed for the generation of both the training dataset and the validation dataset), or to each grid point (for the GHI product map), is used to obtain the SARAH-3 data at the set of locations needed.

Geostationary satellite-based products, such as SARAH-3, also have several challenges associated with the method deriving surface irradiance. Indeed, this method is quite complex: typically, satellite products first characterize the clouds, then assume a clear sky irradiance which involves both climatology data and modeling, and finally they use these data to estimate the surface irradiance \cite{algo_cmsaf}. This involves several steps, and complex calibration and geophysical transfer functions that are based partly on heuristics. As a consequence, even satellite products are not a true, direct observation of the GHI. In particular, it is well known that these products, despite continuous improvements and the development of novel algorithms for the SARAH-3 product version, still have some level of difficulty distinguishing low-altitude clouds from snow or ice cover \cite{skartveit1998hourly,hinkelman2009surface,durr2010verification,castelli2014heliomont,muller2022remote}, and that they also struggle in foggy conditions \cite{algo_cmsaf}. Such challenging conditions, specifically the presence of snow and ice covers, are particularly common and relevant in Norway, given the local weather and climate conditions. As a consequence, we expect it to be possible to improve upon SARAH-3, even though it is already a high-quality product for solar irradiance, by applying ML-based post-processing. Moreover, effects due to the high viewing angle at the latitude of continental Norway, as well as the extent of the satellite pixels footprint, can also contribute to added errors and uncertainties in the SARAH-3 data.

\subsection{Ground truth observations from pyranometers}

In addition to the predictor data from both models and satellite observations described above, we also use in-situ data from pyranometers as a ground truth to train and validate our models. Surface measurements of hourly global horizontal irradiance were collected at 106 locations in Norway between 2016 and 2020. This time period was selected due to good data coverage (see \cite{QAReport}). Earlier than 2016 much less measuring stations were available in Norway. Therefore, this is the 5-year period that is used for training and validation of our ML models. Most of these data are openly available through the Frost API of MET Norway at \url{https://frost.met.no}. These have also been supplemented with additional data from four locations collected in the course of this project.

Since the pyranometer data had not undergone prior quality control, visual inspection and quality flagging routines have been implemented to only use high-quality station and measurement data to train and validate our ML models \cite{QAReport}. More specifically, stations with insufficient data coverage were excluded first, leaving 69 stations that met the criteria of at least 80\% overall coverage and 60\% average coverage for any month of the year in 2016--2020. Then, the data were inspected using both automatic and visual techniques such as time series analysis and comparisons with clear-sky GHI data to identify spurious trends and other significant shortcomings in any station data. As a result, 22 stations were rejected due to issues such as shadowing effects or data inconsistencies, leaving 47 stations in the final dataset. The data from the 47 remaining stations have been flagged for exceedances of theoretical limits, temporal variability, and data distributions \cite{QAReport}. Only valid data have been kept in the final training and validation dataset and the hourly GHI data have been collected in a netCDF file. The dataset is freely available at \url{https://zenodo.org/records/8082726} \cite{ZenodoData}, together with visualization and quality control plots, scripts, and files that include flags and uncleaned data, as well as the original raw data of the 106 stations.

\begin{figure*}[h]
    \centering
    \includegraphics[width=0.55\textwidth]{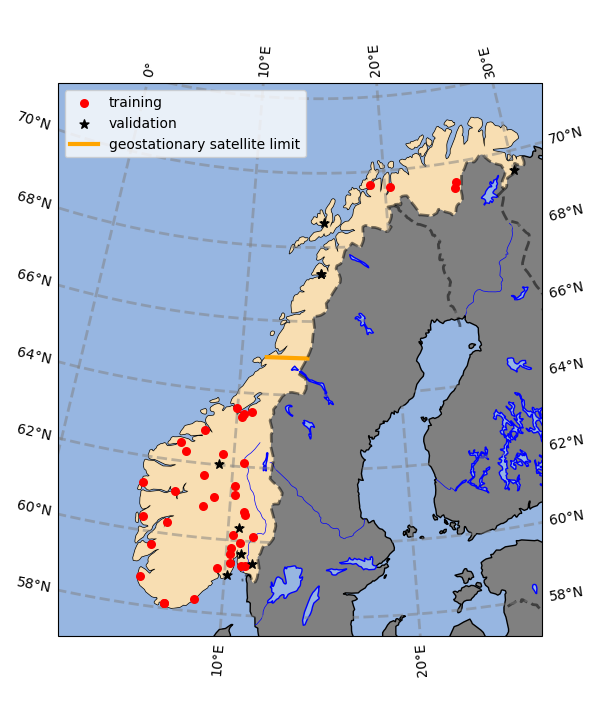}
    \caption{Map of the in-situ pyranometer stations selected to provide the ground truth data (red dots) used to perform model training and stations for validation (black stars).}
    \label{fig:pyranometers}
\end{figure*}

When training and validation of our ML models are performed, we split our data between two disjoint sets of stations used for each purpose, so that we can detect (and avoid) any overfitting. The split between the stations used for training and validation is illustrated in Fig. \ref{fig:pyranometers}. We use 8 stations for validation purposes. The split between stations used for training versus validation is performed so as to represent the general range of conditions over Norway as well as possible in both subsets, based on the general topography and climate conditions observed over Norway.

{\color{black}As visible in Fig. \ref{fig:pyranometers}, there are in situ pyranometer stations in the southern and northern parts of continental Norway. Moreover, the different geographic and climatic areas of continental Norway are well covered: there are stations both on the coast, in the plains of southern Norway, and in the mountain ranges that run along the country. Although there is an area between approximately 64\degree{} N and 67\degree{} N, where unfortunately no station data of sufficient quality is present, there is nothing particular about the geography and climatic conditions in this area relative to those immediately north and south of it. Therefore, we expect that the models trained in this work are still able to perform well in this area, since they have seen conditions similar to those encountered there in their training dataset. This, combined with the fact that simple and robust models are used, as will be described in the next section, mitigates the lack of in situ ground truth data in this specific area. We hope that studies in a few years will be able to remediate this limitation, as more pyranometers are now being deployed over Norway; however, at present, we can only use the data available to us. Since, as will be described in more detail in the next sections, the models rely on pyranometer data only for their training and validation, we are still able to generate GHI data over the corresponding area. Naturally, we expect that having more and better-distributed stations in the future will enable to develop better, larger, and more expressive correction models before overfitting happens, which will allow even better and more accurate ML-corrected maps to be developed.}

\section{Models definition and methodology}

Different ML model configurations have been explored, using linear regression (LR) and neural networks (NNs), finally selecting the best performing model configuration for each point in space and time when producing the final GHI map.

\subsection{Linear regression}

The first model kind we use is a simple linear regression model, taking as input all the parameters presented within Table \ref{tab:model_predictors}. LR models are well known and can be seen as the simplest form of machine learning models \cite{su2012linear,james2023linear}. Despite their simplicity, LR models are known to be effective in many real-world applications \cite{weisberg2005applied}, in particular when several estimators with different and separate error sources (i.e., estimators with non-fully correlated errors) are combined into a Best Linear Unbiased Estimator (BLUE)~\cite{moser1996linear,weisberg2005applied}. This is naturally the case in the present study, as the ERA5 and NORA3 numerical models have quite different setups, and the satellite data provide a separate estimate that is derived from observations.

In the present study, our LR model is simply obtained by applying the $LinearRegression$ function from the sklearn software package \cite{pedregosa2011scikit} to our set of predictors using the ordinary least square optimizer, while selecting pyranometer data as target. While we have investigated the use of more sophisticated optimizers, e.g. ridge regression (see e.g. \cite{segerstedt1992ordinary,mcdonald2009ridge}; ridge regression works by imposing a penalty on the regression coefficient magnitudes using a L2 norm), we found during testing that the plain ordinary least-square optimizer is providing the best results on our task, and it is hence used in the following.

\subsection{Neural Network}

The second kind of models that we use is a fully connected neural network (FCNN). Neural networks have been known for a long time (for example, their property of universal approximators was already discussed in the 1980s~\cite{hornik1989multilayer,cybenko1989approximation}). However, their massive popularity increase happened arguably only recently, concomitantly with the drastic rise in dataset size and available computational power that we have witnessed since the early 2010s \cite{lecun2015deep}. Our motivation for using a simple FCNN is that these models are effective and flexible non-linear function approximators that are supported by a wide range of software packages and easy to train. FCNNs do not require specific prior knowledge about the problem being considered. While large scale models are attracting most attention over the last couple of years (e.g., large image analysis networks \cite{wang2017residual,hurtik2022poly}, the large language models (LLMs) revolution \cite{zhao2023survey,teubner2023welcome}, diffusion models \cite{sohl2015deep,yang2023diffusion}, large reinforcement learning models \cite{silver2016mastering}, or large data-driven global weather models \cite{bi2023accurate,lam2022graphcast}), even simple FCNNs of modest size and complexity are known to be effective nonlinear function approximators that can help on tasks such as error correction \cite{poulton2001computational,krasnopolsky2003some,tedesco2024bias} or data fusion \cite{guo2022information}, given that enough training data are available to perform a representative training.

As simple neural network theory, including the theory describing FCNNs, is now well known across many fields and covered by a very large number of books (e.g., \cite{lecun2015deep}) and reviews (e.g., \cite{abiodun2018state}), we will not reproduce boilerplate materials that discuss simple neural network theory here. Instead, in the following, we consider that the reader has a basic knowledge of the general technicalities of NNs and FCNNs, and we refer the reader without such a background to, for example, \cite{lecun2015deep}. For all practical matters in the present study, FCNN models can be thought of as simple nonlinear function approximators, that can be effectively tuned through gradient descent on the training set to represent the mapping from the combination of model and satellite predictors to the target pyranometer GHI measurements.

We expect FCNN models to work similarly to the LR model described above, except that they have more freedom in designing the mapping function from the predictors to the model output, which can be expected a priori to translate into better accuracy. Indeed, the FCNN can include non-linearity into its mapping, which is not possible for the LR model. Given the relatively moderate amount of data available in the present study, and the fact that both numerical models and satellite observations already provide estimators of the GHI target that have been tuned and optimized by the geoscience community over several decades, we expect moderate accuracy improvements to be obtained with the LR and FCNN models relative to the set of GHI predictors. The improvements obtained by the FCNN model will be the result of both the BLUE effect (similar to the LR model), and possibly the compensation of minor remaining non-linear biases and distorted couplings in the physics of the models and geophysical transfer function of the satellite system, if these exist.

From a practical implementation point of view, neural network methods are known to be potentially sensitive to class imbalance \cite{zhou2005training}. Therefore, since over half the raw training data correspond to night or very low light conditions, we exclude samples for which the GHI is lower than $P_{min}$=50 W/m$^2$ (which is typically around 5\% of the mid-day summer value) from our training and validation dataset. This means that the FCNN that we train can only be used when the incoming solar radiation is equal or higher to $P_{min}$. In other cases, the LR model will be used.

We have tested several different FCNN architectures, optimizers, loss functions, and metaparameters in the early phases of the training process. We found that a simple FCNN with 4 layers of 60 neurons each, using the rectified linear unit (ReLu) activation function, the mean absolute error loss, and the Adagrad optimizer, gives the best validation loss over the present task. This results in a small network with slightly more than 13k trainable parameters in total. We use non-trainable linear scaling normalization layers to scale the input and output to the FCNN, so that the input and output of the hidden neuron layers, as well the neuron activation values inside the FCNN, are typically scaled to the range [-1,1]. The network and its training and prediction routines are implemented in Tensorflow/Keras \cite{tensorflow2015-whitepaper,chollet2015keras}. We find through trial and error that an initial learning rate of 2e-4 (which is then tuned by the Adagrad optimizer as learning progresses) results in satisfactory training within around 2000 epochs. These parameters are typical of what is expected for a small and simple FCNN as implemented here. We have also tried introducing more complexity in the FCNN setup in exploration phases of the training process, such as inserting dropout layers, adding a regularization term to the loss function, inserting batch normalization layers, and using more and larger layers, but this did not provide meaningful improvement, and sometimes even degraded (due to overfitting), the validation loss. This is typical of training processes that correspond to a relatively simple task, for which a moderate amount of data are available, so that adding too much additional complexity in the FCNN does not help improving the validation loss (and can even lead to overfitting). This is consistent with our experience on previous similar projects where moderate amounts of in-situ data were available, and models were already providing a reasonably good prediction of the phenomena considered, so that the FCNN only needs to add small changes on top of the predictors already provided \cite{tedesco2024bias}.

\subsection{GHI map production}

The main operational goal of this work is to generate a ML-optimized GHI map that covers continental Norway over a 30-year period. For this, we need to be able to produce an estimate of the GHI even in cases where a specific input product is unavailable (because it is not available over the whole domain, as is the case with the SARAH-3 data, or because there were some operational production issues for a specific time and place). As a consequence, we train several flavors of both the LR and FCNN models, using different sets of input data. These are summarized in Table \ref{tab:model_summaries}. We then produce GHI estimates using each of the models that can be run (i.e., for which the predictors required are available) at each point in space and time, and we select the best (i.e., the one with the highest model\_index following the validation presented below) that can be successfully produced as the corresponding best estimate for the GHI. Hence, if all data sources are available at a given location and time, the best available model (i.e. the FCNN model using the predictors from all the data sources) is used to estimate the corresponding solar irradiance. If some data are not available for a given point in space or time to apply the best ML model, a simpler model is used instead for the corresponding time and location estimate.

{\color{black}Note that more models could be built and trained, using different subsets of predictors in addition to what is presented in Table \ref{tab:model_summaries}. For example, one could train both LR and FCNN models that use only the ERA5 or only the NORA3 predictors (effectively creating 4 more models, LR\_ERA5, FCNN\_ERA5, LR\_NORA3, FCNN\_NORA3). Due to how the present project developed, and owing to the more modest gains that can be obtained from applying model correction to a single category of predictors at a time (see Appendix C), this is not implemented in the final GHI product that we provide. Moreover, since ERA5 and NORA3 are reanalysis, both these model data are available with good quality for the vast majority of samples in time and space, so that it is seldom that the model\_index 3, or a fortiori 2, cannot be produced. Therefore, in our context, leveraging additional LR and FCNN models based on NORA3 or ERA5 alone would provide a limited improvement only (this is documented in more detail in Appendix C). Although such models could also be used in the future during the generation of the GHI map, this is not the case here.}

\begin{table*}[]
    \centering
    \begin{tabular}{|c|c|}
        \hline
        model\_index & Model description \\
        \hline
        \hline
        6 & FCNN: the best ML model, a simple FCNN, using all data sources \\
        \hline
        5 & LR: a simple LR model, using all data sources \\
        \hline
        4 & GHI\_SARAH-3: the SARAH-3 data source alone \\
        \hline
        3 & FCNN\_NOSAT: a ML FCNN model, using only model (ERA5 + NORA3) data \\
        \hline
        2 & LR\_NOSAT: a simple LR model, using only model (ERA5 + NORA3) data \\
        \hline
        1 & GHI\_ERA5: the ERA5 data source alone \\
        \hline
        0 & GHI\_NORA3: the NORA3 data source alone \\
        \hline
    \end{tabular}
    \caption{The summary of LR and FCNN model flavors used to generate the final 30-year GHI map. At each location in space and time, the best (i.e., highest model index) valid prediction for which all data sources are available is used. The model\_index field is included in the final data product released to the end users.}
    \label{tab:model_summaries}
\end{table*}

Note that the pyranometer data used in the present study are only covering continental Norway, over land. Therefore, sea areas and other land locations, such as Svalbard, are not covered by the final GHI map product, since the ML models are not trained and validated over these areas. This is enforced by masking out the areas where the ML models are not applicable in the final GHI product. As mentioned previously, the SARAH-3 data are not available further than 65\degree{} north, since this data source is based on geostationary observations that cannot be made close to the poles. Therefore, only model indexes 3 and lower are obtained in the northernmost part of Norway.

The full input FCNN model (model\_index 6) relies on all data sources, and is the most sensitive to the input source data quality. {\color{black}More generally, FCNN models, being (non-linear) neural networks, are potentially the models that are the most sensitive to unusual or previously unseen (during training) input data. Therefore, we prefer not to risk extrapolating the FCNN models outside of the strict range of inputs for which they were trained. By contrast, the LR models, being simple linear models, are more robust to inputs that are slightly unusual. As a consequence, we take extra precautions with regard to the FCNN models. More specifically, we check that i) the incoming solar radiation is greater than 50 W/m$^2$ (since we set a lowest value cutoff of 50 W/m$^2$ to filter the training data to avoid class imbalance due to the night, as highlighter higher up), ii) all the input variables are valid within the range seen during training, before accepting that FCNN models can be used. This may not be the case if, e.g., strong precipitations happen that temporarily reduce the amount of solar radiation, or lead to extreme values on some of the predictors to the FCNN models -- in which case, the LR models will be used instead. This is also not the case in some specific locations, such as glaciers in the Norwegian mountains.}

As a consequence, GHI predictions with model\_index 6 are only available south of 65\degree{} north (i.e., where geostationary satellite data are available), when all data sources are of high quality and when the GHI is higher or equal to 50 W/m$^2$. In all other cases, predictions of lower model\_index are provided (in which case, we provide the output corresponding to the highest model\_index model that could be produced given available data). Finally, due to challenges with some of the data sources over the period 2012-05 to 2012-08, only ERA5 data (corresponding to model\_index 1) are used on the corresponding period North of 65\degree{} N.

\section{Results, discussion, and GHI product}

In the following, we compare and analyze our results based on the different available model configurations, before describing the final GHI map product in detail.

\subsection{Analysis and comparison of models performance}

The key metrics (computed on the validation stations) comparing the most important models are summarized in Table \ref{tab:compare_models}. The different columns indicate the values obtained from observations (i.e., the ground truth pyranometer data), GHI estimate fields available in numerical models (NORA3 and ERA5) and in satellite data (SARAH-3), as well as from ML models. ML models with the \_NOSAT postfix (i.e., LR\_NOSAT and FCNN\_NOSAT) indicate models flavors that do not use the SARAH-3 data as predictors, and these are trained and evaluated over all of Norway. ML models without this postfix (i.e., LR and FCNN) indicate model flavors that use SARAH-3 data as predictors, and these models are trained and evaluated using only the data available south of 65\degree{} north, where SARAH-3 data are available. {\color{black}This is reflected in the number of samples used to compute each statistics, which is also reported in Table \ref{tab:compare_models}.}

The results presented in Table \ref{tab:compare_models} are in good agreement with what can be expected based on the characteristics of each product or model. The two numerical models (NORA3 and ERA5) have the highest level of both mean absolute error (MAE) and root mean square error (RMSE) relative to observations. NORA3 has slightly higher error levels compared to ERA5, but has also significantly higher and more realistic levels of variability in the predicted data (indicated by the "std. of data" column). This corresponds well to the fact that NORA3 is a higher resolution model compared to ERA5. Said otherwise, ERA5 has a tendency to spatially smooth out physical processes and the predictions that it provides, which produces slightly better error metrics at the cost of reduced intrinsic variability in its output. The SARAH-3 satellite data have better MAE, RMSE, and variability levels compared to the data from numerical models. Naturally, the SARAH-3 product still has non-negligible levels of MAE and RMSE, as well as slightly lower intrinsic variability, compared to the actual observations. This is a natural consequence of the finite footprint of the satellite data acquisition pixels, as well as of the post-processing that is necessary to derive an actual GHI observation from the physical signals acquired by the satellite, and a variety of error sources induced by high angle of incidence at the latitude considered. As noted earlier, snow and ice cover also add to the uncertainty of the satellite-derived irradiance. 

Regarding the ML models, the FCNN models have slightly better performance compared with the LR models, both when satellite data are used and when they are not. Both the LR and FCNN models using all data inputs (including SARAH-3) have lower levels of MAE and RMSE compared to the SARAH-3 data. In the case of the LR model, this comes at the cost of a non-negligible reduction in the intrinsic variability of the predicted GHI, while the FCNN model has an intrinsic variability that is closer to that of SARAH-3 and the pyranometer observations. This can naturally be understood by the "purely averaging" effect obtained applying a linear regression model, compared to the non-linear properties of neural network based techniques. ML models that do not use SARAH-3 satellite data as an input (i.e., LR\_NOSAT and FCNN\_NOSAT) have, unsurprisingly, higher levels of MAE, RMSE, and slightly lower levels of intrinsic variability. This is understandable, as the SARAH-3 data input was the one that brought the highest accuracy GHI estimates, with the most realistic level of data variability. However, these \_NOSAT models still significantly outperform the NORA3 and ERA5 GHI estimates by themselves, which indicates that they manage to extract meaningful information from the numerical model predictors provided, and to use this to improve on the GHI predictions.

\begin{table*}
    \centering
    \begin{adjustwidth}{-.35in}{-.35in}
    \begin{tabular}{|c|c|c|c|c|c|c|c|c|c|}
        \hline
         & model index & 0 & 1 & 4 & 5 & 6 & 2 & 3 \\
        \hline
        (W/m$^2$) & obs. & GHI\_NORA3 & GHI\_ERA5  & GHI\_SARAH-3 & LR & FCNN & LR\_NOSAT & FCNN\_NOSAT \\
        \hline
        \hline
        MAE vs. obs. & N.A. & 78 & 72 & 59 & 54 & 46 & 67 & 62 \\
        \hline
        RMSE vs. obs. & N.A & 116 & 103 & 87 & 78 & 69 & 94 & 92 \\
        \hline
        std. of data & 215 & 207 & 190 & 209 & 197 & 204 & 190 & 201 \\
        \hline
        bias vs. obs. & N.A & 27 & 7 & -7 & 3 & 3 & 3 & 6 \\
        \hline
        {\color{black}nbr. samples} & {\color{black}119k} & {\color{black}119k} & {\color{black}119k} & {\color{black}78k} & {\color{black}78k} & {\color{black}78k} & {\color{black}119k} & {\color{black}119k} \\
        \hline
    \end{tabular}
    \end{adjustwidth}
    \caption{Summary of i) the error (mean absolute error (MAE) and root mean square error (RMSE)) of the different GHI predictors and ML models relatively to the in-situ pyranometer observations from the validation stations, ii) the intrinsic variability of the data for each GHI estimate ('std. of data'), and iii) the bias relative to observations ('bias vs. obs'). {\color{black}We also report the number of samples ('nbr. samples') on which each statistic is computed (rounded to the nearest 1k), which varies since we have a total of 8 validation stations, but only 5 of these are in the area that is covered by satellite data.} model\_index refers to definitions in Table 2. All values reported are computed over the set of validation stations.}
    \label{tab:compare_models}
\end{table*}

\begin{figure*}[h]
    \centering
    \includegraphics[width=0.99\textwidth]{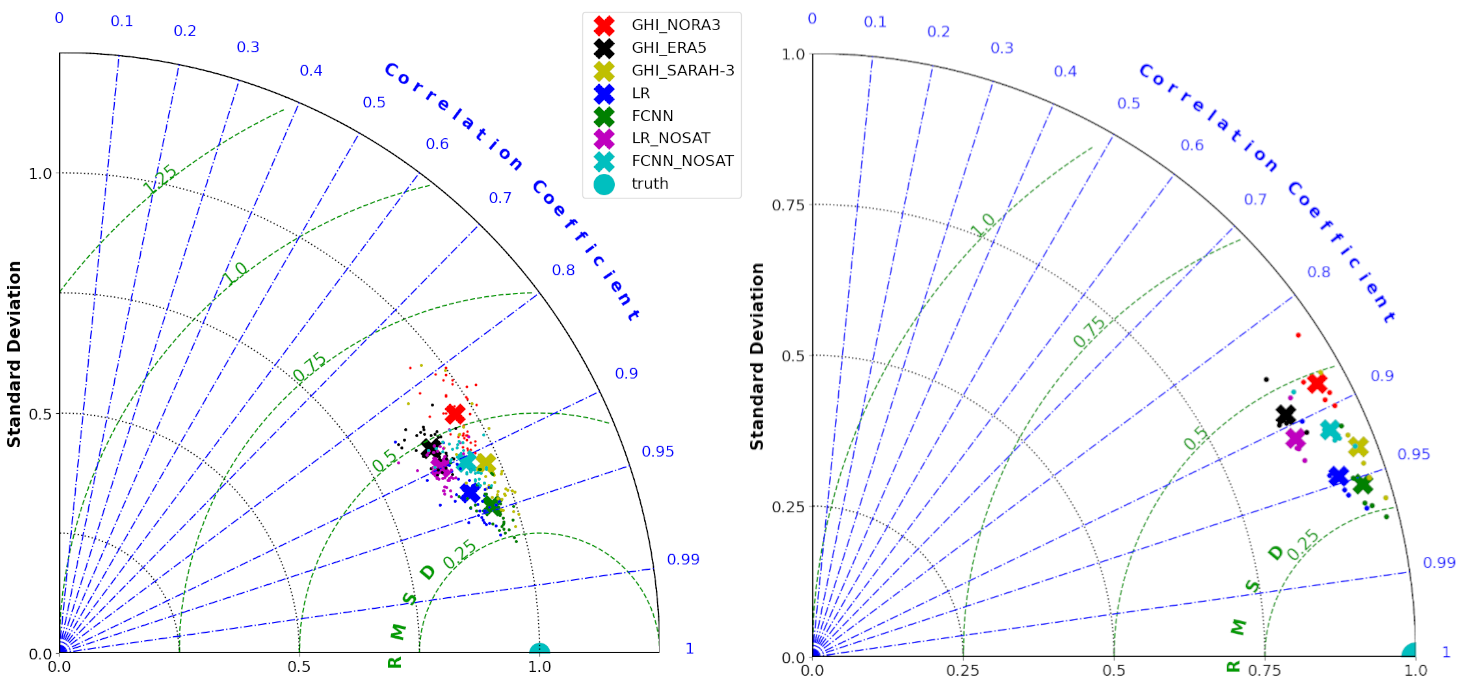}
    \caption{Taylor plots summarizing the performance of the different predictors and models. The acronyms used correspond to Table \ref{tab:model_summaries}. GHI\_ estimates are obtained directly from the corresponding data sources. LR indicates a linear regression model, and FCNN a neural network model. Models with the \_NOSAT postfix do not use the SARAH-3 satellite data source, in contrast to models without the postfix. Left: performance estimated on the training stations. Right: performance estimated on the validation stations. This confirms the general picture provided by Table \ref{tab:compare_models}, as well as confirms that there is no noticeable overfitting.}
    \label{fig:taylor_plots}
\end{figure*}

The data presented in Table \ref{tab:compare_models} can be extended and shown in Taylor plots, as shown in Fig. \ref{fig:taylor_plots}. Taylor plots show the amount of normalized intrinsic variability (radial distance) and the correlation coefficient (angle) between model and ground truth data. These are computed between the normalized ground truth pyranometer data (where normalization is performed individually for each station so that the normalized intrinsic variability reported for each station is matched to 1), and the normalized GHI prediction obtained at each individual station with each model (there, also, the same normalization is used for each station as was used for the normalized pyranometer data). This enables effective visualization of both the accuracy and the amount of intrinsic variability in each model output for each station (small dots), as well as the average performance of each model across each set of stations (thick crosses). The Taylor plots are shown for both the training (left panel) and the validation (right panel) subsets of stations.

The general pattern for models location in the Taylor plots in Fig. \ref{fig:taylor_plots} is very similar between the training and validation datasets, with variability for individual stations compared to the average values over each set of stations which comes from the intrinsic variability between stations, as can be expected given the wide range of conditions encountered over Norway (e.g. coastal areas, plains, mountains, and both temperate and sub-polar climate regimes). The very similar localization of the results across the Taylor plots for both the training and validation cases confirms that there is no noticeable overfitting. Indeed, this implies that there is no significant performance degradation observed by applying the models on previously unseen data compared to the training data. Moreover, the results shown in Fig. \ref{fig:taylor_plots} confirm the general impression provided by the scalar metrics in Table \ref{tab:compare_models}. More importantly, the Taylor plots indicate a clear clustering of individual station data (small dots) around individual model cluster centers (thick crosses), confirming that the different predictors and models (different colors) have significantly different accuracy. The GHI estimates obtained directly from numerical models have the lowest coefficients of correlation (though these are already quite high, at over 0.85). ERA5 has higher correlation with pyranometer in-situ observation data than NORA3, though this comes at the cost of too low variability in the data, indicating increased levels of smoothing as pointed out previously. The LR model without satellite data is able to obtain slightly higher correlation than ERA5, but has typically the same low variability. The FCNN model without satellite data provides a bit higher correlation, and variability that is closer to the real data. The SARAH-3 data have higher correlation and better variability than all data sources and all models that do not rely on satellite observations. The LR and FCNN models with satellite information are the best GHI estimates available, with the FCNN being slightly better in both correlation and variability values.

\begin{figure*}[h]
    \centering
    \includegraphics[width=0.9\textwidth]{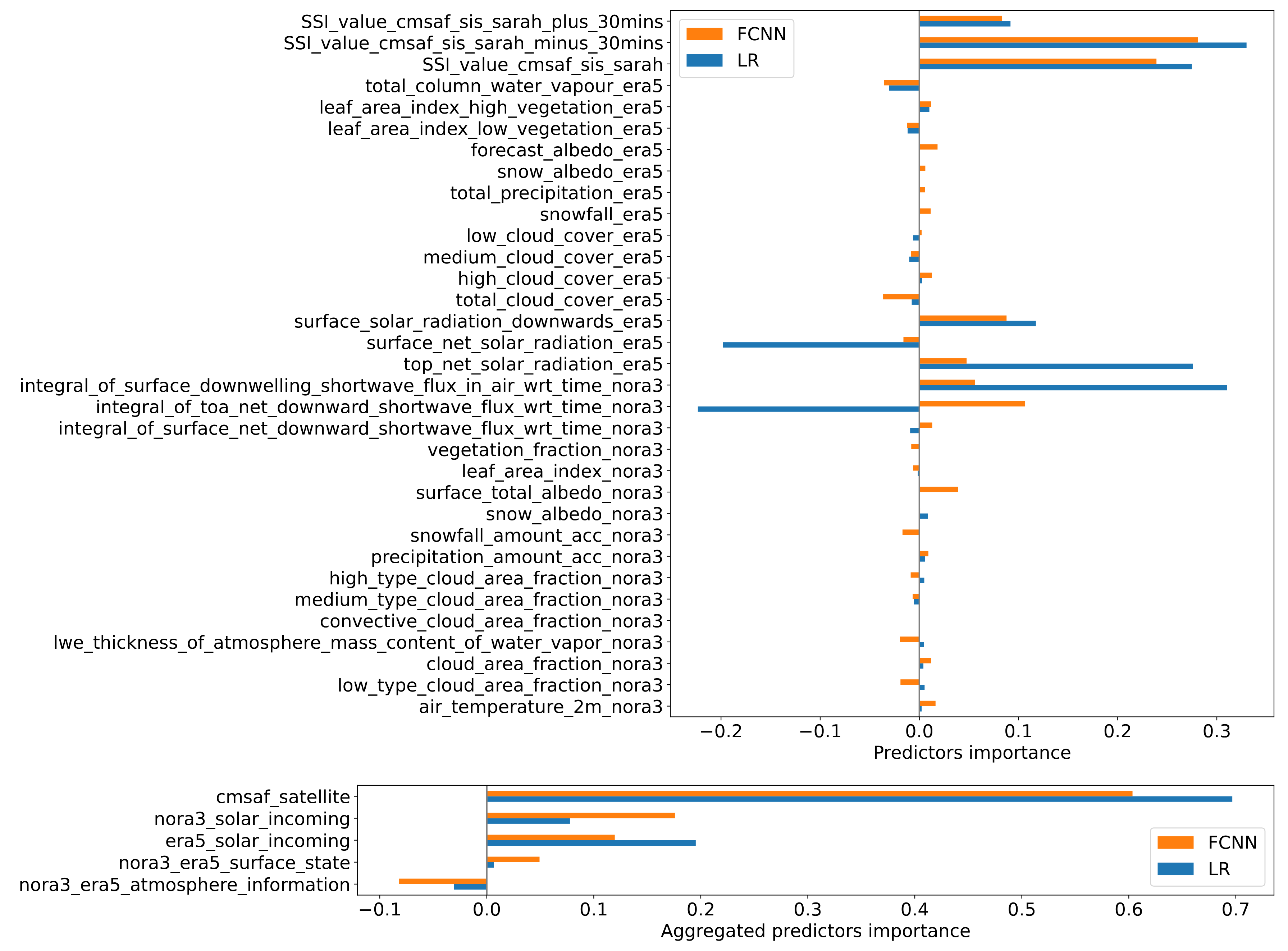}
    \caption{Estimate of the linearized importance of the predictors used by the LR and FCNN models. This shows (in a linearized sense for the FCNN), how much weight is given to the different input predictors. Top: individual overview for all predictors. This can be misleading, with spurious canceling importance given to closely related predictors. Bottom: importance averaged over a few categories. The general importance distribution among categories of importances (bottom) is in good agreement with what would be expected from a best linear unbiased (BLUE) frame of thinking: more importance is given, in a linearized sense, to the highest quality predictors.}
    \label{fig:predictors_importance}
\end{figure*}

All these results confirm that the LR model flavor is able to provide better GHI estimates than any of its input predictor taken in isolation, which is as expected from the BLUE theory highlighted earlier. In addition, the FCNN model, which also includes non-linearity, performs slightly better than the LR model, as could be expected. It is notoriously difficult to interpret FCNNs to try to understand their inner workings, and even the interpretation of simple models such as LR can be misleading~\cite{alqaraawi2020evaluating}. However, we can still attempt to inspect how each model weights the different predictors available, and try to compare how the LR and FCNN models operate. For this, we present a measure of predictors importance in Fig. \ref{fig:predictors_importance}. For the LR model, the predictor importance (predictor\_importance\_LR) is defined as:

\begin{equation*}
\begin{multlined}
    predictor\_importance\_LR = \\
    predictor\_coefficient~\cdot~std(predictor)~/~std(prediction), 
\end{multlined}
\end{equation*}

\noindent where $predictor\_coefficient$ is the LR coefficient associated with the corresponding predictor, $std(predictor)$ is the standard deviation (intrinsic variability) of the corresponding predictor, and $std(prediction)$ is the standard deviation of the target model output, i.e. the GHI observations. Using this definition, one can measure the variance-normalized contribution of each predictor to the model output. As visible in the formula, this normalized importance is a combination of both the value of the linear regression coefficient associated to each predictor, and of the variability of the corresponding predictor. In this definition, the magnitude of a predictor importance indicates how much this predictor contributes to the variance of the output, and the sign indicates if this predictor is correlated (positive) or anticorrelated (negative) with the output.

Computing a predictor importance is more challenging in the case of the FCNN. Since the overall performance of the FCNN is relatively similar (though slightly better) to the one of the LR, we can hypothesize that the FCNN is relatively close to a linear regression model, with some additional non-linearity added on top of the linear LR behavior. Therefore, we expect that the FCNN is relatively close to its linearized approximation at each point in the prediction space. As a consequence, we leverage the local linear sensitivity of the FCNN relative to each predictor, averaged over the whole dataset, to compute the following predictor\_importance\_FCNN estimate:

\begin{equation*}
\begin{multlined}
    predictor\_importance\_FCNN = \\
    \frac{1}{N} \sum_N \frac{\partial FCNN\_prediction}{\partial predictor}~\cdot~ \\
    std(predictor)~/~std(prediction),
\end{multlined}
\end{equation*}

\noindent where $N$ is the number of samples in the dataset, $predictor$ is the predictor input to the FCNN, and $FCNN\_prediction$ is the output GHI prediction by the FCNN. Since neural networks are fully differentiable computational graphs, the partial derivative term is easily obtained by leveraging the functionality natively built into Tensorflow/Keras. Our approach is relatively similar to the saliency map and the Grad-CAM methods \footnote{\url{https://keras.io/examples/vision/grad_cam/}}\textsuperscript{,}\footnote{\url{https://christophm.github.io/interpretable-ml-book/pixel-attribution.html}} that are used to investigate which parts of an image contribute the most to the classification score for each class \cite{selvaraju2017grad}. This is carried out by investigating the sensitivity of the prediction performed by the network to variations in the model input. However, in the present case, we can average the sensitivity over all training samples in order to obtain an average sensitivity, which gives an indication of the averaged locally linearized behavior of the FCNN model.

The results for the two best performing models, i.e. the LR and FCNN models using all predictors, are presented in Fig. \ref{fig:predictors_importance}. We present both 1) the importance of each predictor for the LR and FCNN models taken individually (top panel), and 2) the aggregated importance for groups of closely related predictors for the same two models (bottom panel). The reason for presenting both the individual and aggregated predictors importance is that when several predictors are closely related (for example, the surface\_solar\_radiation\_downwards\_era5 and the surface\_net\_solar\_radiation\_era5 predictors are very similar to each other, and differ only due to the effect of relatively minor contributions), the models may have some degree of spurious sensitivity to these predictors. In particular, the models may attribute importances with canceling effects to these closely related predictors. Hence, the true importance of these predictors taken as a whole may look different from their individual (canceling) importances, as visible in Fig. \ref{fig:predictors_importance}.

The general results presented in Fig. \ref{fig:predictors_importance} confirm our previous impressions about the behavior of the models. Typically, more importance is given to the SARAH-3 predictors, which are also the highest quality ones. This is consistent with what would be expected from the BLUE theory. In addition, some importance is given to both the NORA3 and ERA5 predictors. Interestingly, non-negligible importance is given to the surface state predictors. This may indicate (though this is speculative as it is challenging to interpret the weights and sensitivities of a NN or even a LR model) that this information is used to correct, e.g., biases and errors in SARAH-3 measurements due to snow and ice cover. Generally, Fig. \ref{fig:predictors_importance} confirms our previous interpretation that the FCNN model is relatively close to the LR model, with some additional non-linear modulation taking place that explains for the slightly better performance. However, we also want to stress that one should not over-interpret such predictor importance map, as similar techniques used in image analysis and that have inspired our present approach, such as Grad-CAM \cite{selvaraju2017grad} and saliency maps \cite{hsu2023explainable}, are notorious for being difficult to interpret into details and are known to be potentially misleading.

\subsection{GHI map product}

We generate the final product map by applying the models described in Table \ref{tab:model_summaries} over continental Norway with a time resolution of 1 hour and a spatial resolution of 0.03\degree{} (approximately 3.3 km), covering the 30-year period 1991--2020 (from the start of January 1st, 1991, to the end of December 31st, 2020). {\color{black}In order to do so, we interpolate (using the nearest neighbor method) the input data used by the ML models to the grid on which the final GHI map is produced. The choice of a 0.03\degree{} resolution for the final product is motivated by i) the size on disk of the final map product, and ii) the resolution of the input data. Regarding i), the final map is about 45~GB is size at the current resolution. This is a manageable amount of data that allows cheap hosting, downloading a local copy of the data if necessary, fast access and slicing, and is generally convenient to work with. Regarding ii), ERA5 has a resolution of 31~km, NORA3 3~km, and SARAH-3 0.05\degree{}. Therefore, the input data with highest resolution, NORA3, typically has a resolution slightly better than 0.03\degree{}. Consequently, it would not make sense to generate a GHI map using our methodology while “artificially” interpolating to a resolution much higher than that of the finest data input. Naturally, it is still an interesting question to generate even higher resolution maps, but this would rather be the topic of a project focusing on generation of super-resolution data, which may require additional and different methodology compared to what we use here.}

Links to the product files are provided in Appendix A. If all data sources are available at a given location and time, the best available model (i.e., model\_index 6) is used to estimate the corresponding solar irradiance. If some data are not available for a given point in space or time to apply the best model, a simpler model (the one with highest model\_index that can be applied to the case) is used instead for the corresponding time and location estimate. For each location and time, the best (i.e., a higher "model index") data output that is available is used in the final product, and these data are collected together with the corresponding model\_index. The data files of the product are hosted as netCDF-CF files on the THREDDS server of MET Norway: \url{https://thredds.met.no/thredds/catalog/sunpoint/ML-Optimized-Maps/catalog.html}.

More technical details are provided in Appendix B. In particular, we want to stress that the data are provided following 2 different data-chunking layouts, and that choosing the right layout is critical for data retrieval performance.

\begin{figure*}[h]
    \centering
    \includegraphics[width=0.99\textwidth]{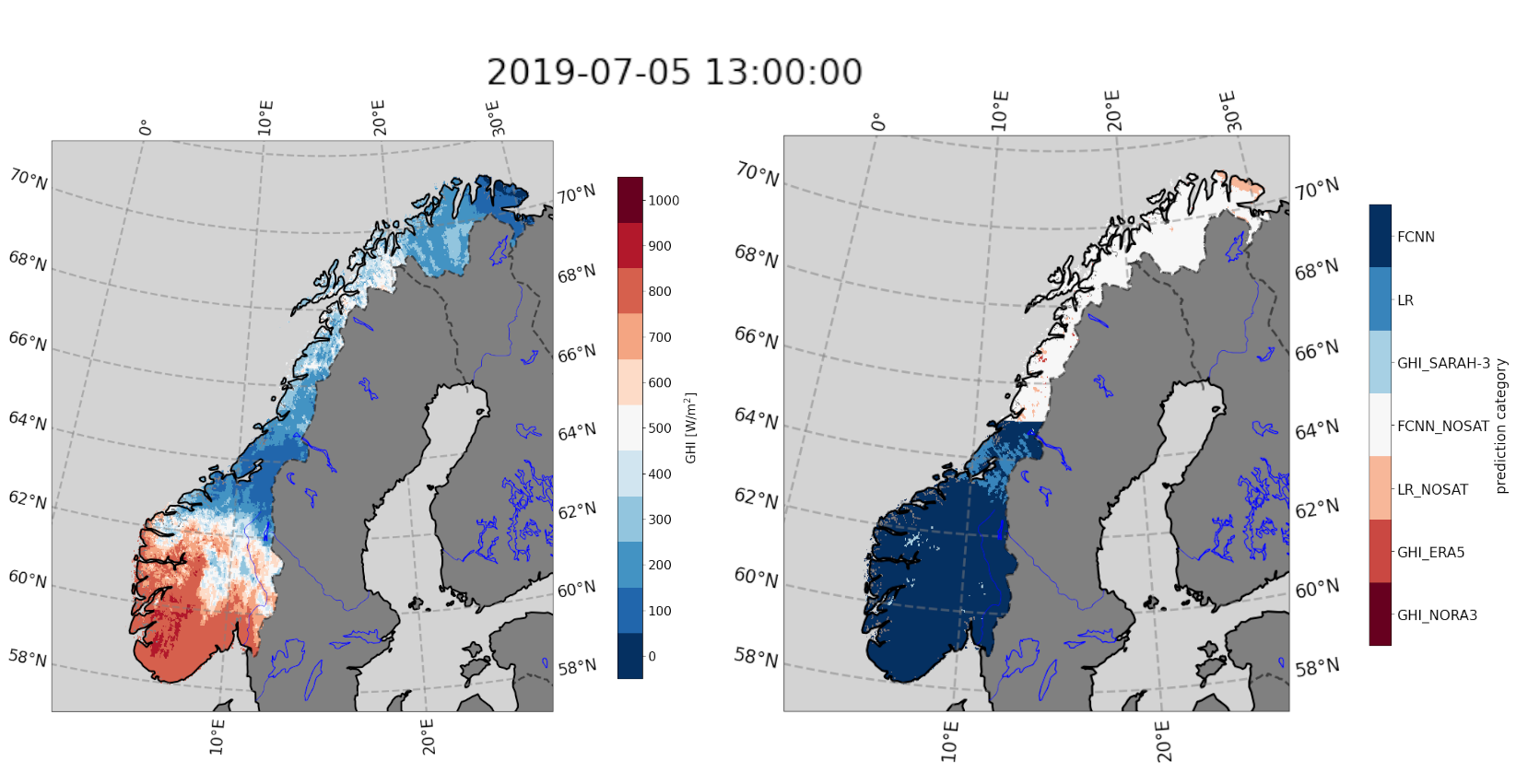}
    \caption{Illustration of the final GHI product. Snapshot for one point in time (2019-07-05 13:00 UTC) for the instantaneous GHI (W/m$^2$) prediction over Norway (left panel) and the associated model\_index values (right panel).}
    \label{fig:instantaneous_maps}
\end{figure*}

\begin{figure*}[h]
    \centering
    \includegraphics[width=0.5\textwidth]{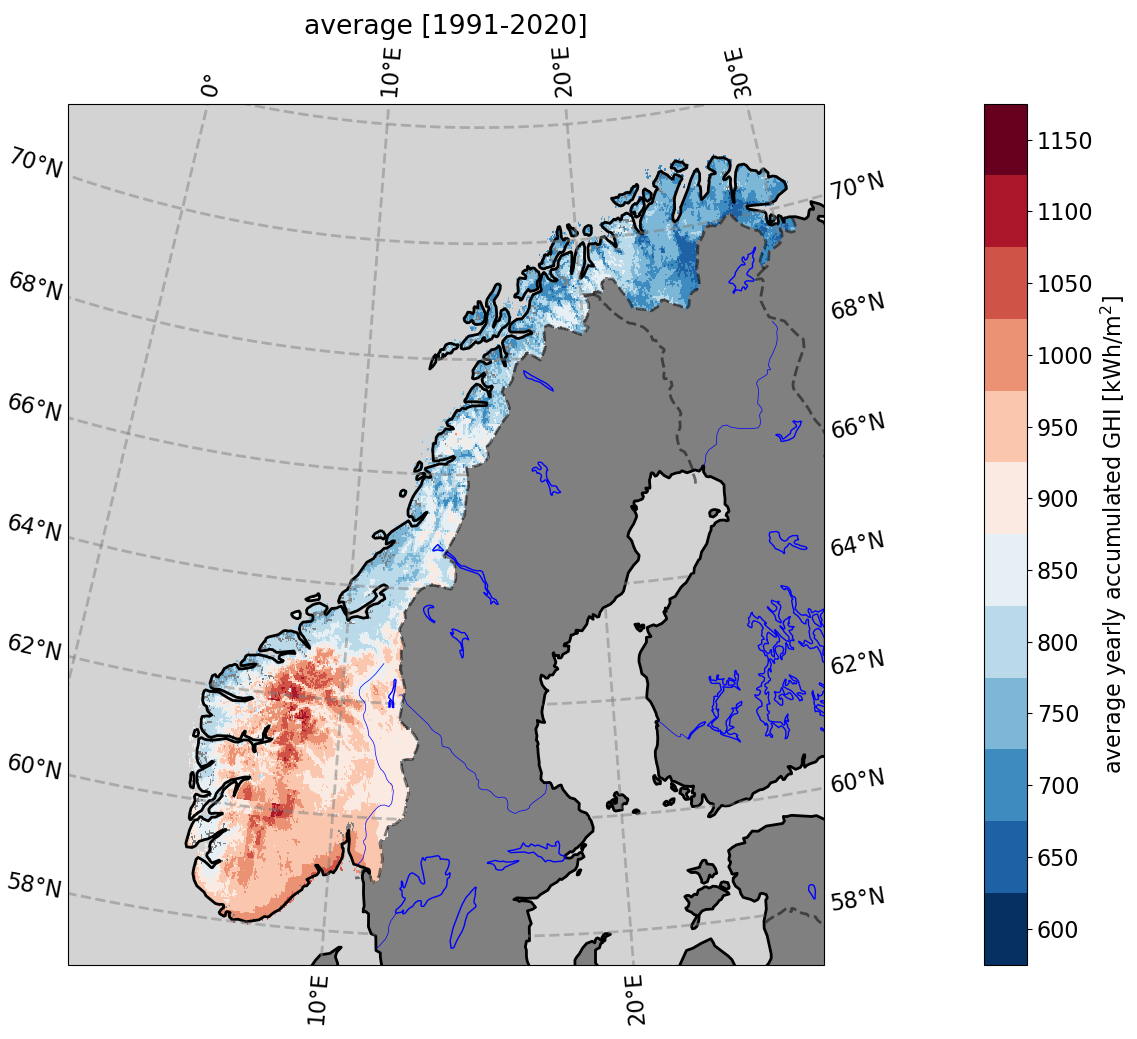}
    \caption{Average of the yearly accumulated GHI over Norway (kWh/m$^2$) for the period 1991-2020.}
    \label{fig:average_5years}
\end{figure*}

A typical map of the predicted GHI (left panel) and the model\_index used at each point (right panel) for a specific point in time is presented in Fig. \ref{fig:instantaneous_maps}. We see that the FCNN model with SARAH-3 input is used on most of the southern parts of Norway, as expected. Concomitantly, the FCNN model without SARAH-3 input is used on the northern part of Norway. There are a few locations where either the GHI is too low to use the FCNN models, or some predictors are {\color{black}taking unusual values due to specific weather conditions at the time of the snapshot that is presented, and other methods, in particular the LR models output, are used over the corresponding areas. This is especially visible in the area around 64N, where heavy cloud coverage is present which locally reduces the GHI for the specific date and time shown.} A map of the averaged accumulated yearly GHI (kWh/m$^2$) for the whole period 1991-2020 is presented in Fig. \ref{fig:average_5years}. Values above 1000 kWh/m$^2$ are found along the coast of southeastern Norway and in the high mountain region of southern Norway. Lower annual values are seen in the west coast and in central Norway (800-850 kWh/m$^2$) dropping to 700-750 kWh/m$^2$ in northern Norway. It is also noted that higher annual values are encountered in the mountainous areas in northern Norway. The values obtained in southern Norway are in line with what is reported by other products, e.g. the GSA shown in Fig. \ref{fig:GlobalSolarAtlas}. However, with our product, data are available over all of Norway (also where there are no data in Fig. \ref{fig:GlobalSolarAtlas} due to the unavailability of geostationary satellite data). There is significant spatial variability in the mean accumulated yearly GHI over Norway. The general variations observed correspond well with physics-based intuition and expectations, with regions in Northern Norway receiving less accumulated solar energy because of the higher latitude and the associated climate. Mountainous areas in Southern Norway are receiving the highest amount of accumulated solar energy in late spring/early summer, likely due to altitude effects (high clear-sky radiation because of few aerosols and low water vapor), high insolation due to low cloud-cover and high albedo values from snow (reflection). Similarly, the southeastern coastal area is characterized by many days with clear skies or little cloud cover in summer, giving rise to high annual average surface irradiance.

Finally, we compare our final GHI ML product with in-situ pyranometer data and other products at the validation stations. This is illustrated in Fig. \ref{fig:example_comparisons}. We see, similarly to what has been presented in Fig. \ref{fig:taylor_plots}, that the GHI ML product compares well with observations. The ML product is able, despite the inherent variability associated with individual stations and the use of a combination of the different model\_index predictions at each location, to reduce biases in average compared with other data sources taken individually.

\section{Conclusion and future work}

In the present work, we demonstrate that a simple Machine Learning (ML) data fusion methodology can be used to generate Global Horizontal Irradiance (GHI) maps with improved accuracy compared to any individual input product. The methodology developed here is simple and robust, and could easily be applied to any region of the world where both in situ pyranometer ground-truth point measurements are available to train the models, and several numerical and observational gridded data sources can be used as predictors for estimating the GHI. We expect that different regions of the world have different model biases and mismatches relative to observations and, hence, that specifically trained ML models will likely be needed over different geographic areas.

We leverage this methodology to generate a Machine Learning (ML)-enhanced Global Horizontal Irradiance (GHI) map over Norway for the period 1991--2020 at 0.03\degree{} (approximately 3.3 km) spatial and 1-hour temporal resolution. For this, we collect data from pyranometers deployed over Norway to be used as our ground truth target, and we select data from 2 numerical reanalysis products (ERA5 and NORA3) and 1 satellite data product (SARAH-3) to be used as predictors for our models. The selection of these products is driven by the requirement that they should be available over the 30-year period considered and that the statistical properties of their errors should be as uniform as possible over this time period. We train both simple linear regression (LR) models and slightly more advanced fully connected neural network (FCNN) models. We show that both the LR and FCNN models provide better accuracy and lower noise levels compared to any input data source taken individually. This is a natural and expected result, explainable by the theory behind the Best Linear Unbiased Estimator (BLUE). By combining different data sources about the same target quantity (here, GHI) that contain different modeling or observation noise sources and error characteristics that are not fully correlated across data sources, one can obtain a better estimate of the target by using either a linear or a non-linear combination of these data sources. Out of the different models considered, we show that the non-linear FCNN model using all inputs as predictors performs best, though even the simpler linear LR model already provides significant improvement and has an overall performance which comes close to the FCNN. The GHI map produced using the FCNN and LR models is made publicly available on the THREDDS server of MET Norway at \url{https://thredds.met.no/thredds/catalog/sunpoint/ML-Optimized-Maps/catalog.html}. We expect that this ML-optimized GHI product will be useful for a variety of purposes, ranging from meteorological and climate studies to the investigation of solar resource potential over Norway.

The present work could be further improved on in the future. In particular, the data used for training and evaluating the model cover the period from 2016 to 2020, i.e. 5 years of data. By now, more years of data are available, which would extend the training dataset. Moreover, a significant number of new pyranometers have been deployed in Norway recently, and the maintenance and calibration of the pyranometers operating in Norway has received increased attention following the growing interest in evaluating possible solar resources. As a consequence, the ground truth data foundation for training GHI models using the methodology presented here has improved since the beginning of this project. Finally, we used only geostationary satellite data in this work, due to their ease of use and, in particular, the ease with which these can be interpolated to a regular grid in time and space. However, geostationary satellite GHI measurements are either unavailable or significantly more uncertain at the high latitudes of Norway and in snowy or icy conditions, compared to lower latitudes. In this regard, it could be useful to also use polar orbit satellite GHI data for future work, though this presents additional technical challenges due to the varying swath coverage and overflight times. This will require further consideration on how these data can be incorporated into the training of models and evaluation functions. Finally, new weather model products will soon be released, such as the ongoing ERA6 re-analysis which will have a resolution nearly twice as fine as ERA5, and will be produced using a model version that includes 8 more years of R\&D, validation, and tuning compared to ERA5. This will provide potential for significantly improved predictors for future ML models as well.

As a final word of caution, we want to emphasize that this product is a best estimate of the GHI over the period from 1991 to 2020. In particular, these estimates of past conditions may not be fully representative of long-term future GHI values if significant climatic variations occur in Norway following, e.g., climate change.

\section*{Acknowledgements}

This document and the GHI maps have been produced in the context of the SUNPOINT project, funded by The Research Council of Norway (grant number 320750).

\section*{Appendix A: open data release and links to the products}

The machine learning-generated irradiance map over Norway is available on the THREDDS server of the Norwegian Meteorological Institute at \url{https://thredds.met.no/thredds/catalog/sunpoint/catalog.html}.

Relevant code examples that indicate how to use the final product are available on GitHub at \url{https://github.com/metno/2024_Sunpoint_solar_irradiance_map_over_Norway_1991-2020}. A short user manual is provided there, at \url{https://github.com/metno/2024_Sunpoint_solar_irradiance_map_over_Norway_1991-2020/blob/main/2024_SunPoint_user_instructions_v1.pdf}.

We especially want to highlight to users that the 1-hour-resolution irradiance map is available in two flavors: (i) field access files (i.e., \url{https://thredds.met.no/thredds/catalog/sunpoint/ML-Optimized-Maps/hourly/field_access/catalog.html}, which are stored (chunked) in a manner to be most effective for retrieving a field at a given point in time), and (ii) point access files (i.e., \url{https://thredds.met.no/thredds/catalog/sunpoint/ML-Optimized-Maps/hourly/point_access/catalog.html}, which are stored in a manner to be most effective at retrieving a timeseries at a given point in space). The actual data contained in both types of files are identical, except for this aspect of the data-chunking layout. The layout is critical for the efficiency of the data retrieval: using the wrong kind of file for a given task (i.e., (i) retrieval of a map at one time versus (ii) retrieval of a timeseries at one time) can result in speed penalty factors up to a factor of 300. See the Jupyter notebook \url{https://github.com/metno/2024_Sunpoint_solar_irradiance_map_over_Norway_1991-2020/blob/main/SunPoint_data_user_example.ipynb} for more details. Moreover, retrieving data through the THREDDS server introduces intrinsic latency and slower performance than using fast (e.g., SSD) local drives containing the same data. Therefore, if many data extraction requests are to be performed by end users and speed is critical, such end users should consider downloading locally a copy of the files exposed on the THREDDS server, and use these local files to perform their data extractions.

These data are released under the Norwegian license for public data (NLOD) and Creative Commons 4.0 BY, see \url{https://www.met.no/en/free-meteorological-data/Licensing-and-crediting}.

\section*{Appendix B: example jupyter notebook of how to use the data}

We provide a Python example of how to access and plot the data in the following Jupyter notebook: \url{https://github.com/metno/2024_Sunpoint_solar_irradiance_map_over_Norway_1991-2020}. This makes heavy use of the xarray package to open netCDF-CF datasets. As highlighted in Appendix A, and as visible in the example notebook, there are 3 types of data files provided as a final product:

\begin{itemize}
    \item (1) solar radiation at ground, chunked to make it efficient to access a full map at a given point in time. These files provide the instantaneous solar irradiance at ground in W/m$^2$ at each point in space and at a 1-hour resolution. Both the best estimate GHI value, as well as the model\_index (see Table \ref{tab:model_summaries}) that was used to produce it, are provided at each point. These files should be used to, e.g., plot the solar irradiance over a given area in space for a single point in time. See the files \url{https://thredds.met.no/thredds/catalog/sunpoint/ML-Optimized-Maps/hourly/field_access/catalog.html} and related.
    \item (2) solar radiation at ground, chunked to make it efficient to access the timeseries for a single point in space over the whole time coverage of the dataset. These are the exact same data as (1) and the units and fields are identical, except that the chunking, i.e. the memory layout of the data, is optimized for a timeseries access pattern. These files should be used, e.g., to extract time series of solar irradiance at a selection of points. See the files \url{https://thredds.met.no/thredds/catalog/sunpoint/ML-Optimized-Maps/hourly/point_access/catalog.html} and related.
    \item (3) monthly accumulated values data files; these files provide the accumulated solar energy in kWh/m$^2$, for each point in space and each month between 1991 and 2020. See the file \url{https://thredds.met.no/thredds/catalog/sunpoint/ML-Optimized-Maps/monthly/catalog.html}.
\end{itemize}

{\color{black}
\section*{Appendix C: additional models (LR\_ERA5, FCNN\_ERA5, LR\_NORA3, FCNN\_NORA3)}

In this Appendix, we summarize the results obtained by applying a LR or a FCNN correction model on top of the set of predictors from ERA5 (leading to models LR\_ERA5 and FCNN\_ERA5) or NORA3 (leading to models LR\_NORA3, FCNN\_NORA3) alone. The methodology used is otherwise similar to what is discussed in the main body of the text, and the sole difference is the selection of the set of inputs to the models, which is reduced to a single reanalysis at a time (though we still use all the predictors corresponding to each of these single models). These models are not used for the generation of the final GHI map, though they could be included in future studies.

The performance of these models on the validation stations is summarized in Table \ref{tab:additional_models}. As visible there and as expected, for each category of input data (ERA5 or NORA3), the FCNN model is slightly better than the LR model, which is itself slightly better than the raw model output. The improvement is less than what is obtained using the FCNN\_NOSAT and LR\_NOSAT models (comparing to the results from Table \ref{tab:compare_models}). This corresponds to expectations, since the gains come from the addition of a correction adapted to the geographic region considered on top of a single model input variables set at a time, while the effect of combining inputs from different models with different intrinsic error sources is not present.

Since ERA5 and NORA3 are reanalysis, the input data from both models are usually available at all points in time and space over the domain considered, so that the FCNN\_NOSAT or LR\_NOSAT models can be run except in a few special cases. Therefore, the fact that we do not use the additional models discussed here in the production of the final GHI product has a negligible impact on the accuracy of the generated GHI map as a whole.
}

\begin{table*}
    \centering
    \begin{tabular}{|c|c|c|c|c|}
        \hline
        (W/m$^2$) & LR\_ERA5 & FCNN\_ERA5  & LR\_NORA3 & FCNN\_NORA3 \\
        \hline
        \hline
        MAE vs. obs. & 70 & 65 & 76 & 67 \\
        \hline
        RMSE vs. obs. & 100 & 97 & 103 & 98 \\
        \hline
        std. of data & 189 & 204 & 183 & 200 \\
        \hline
        bias vs. obs. & 5 & 3 & -5 & 7 \\
        \hline
        nbr. samples & {119k} & {119k} & {119k} & {119k} \\
        \hline
    \end{tabular}
    \caption{{\color{black}Summary of i) the error (mean absolute error (MAE) and root mean square error (RMSE)) of the additional ML models relatively to the in-situ pyranometer observations from the validation stations, ii) the intrinsic variability of each of these data, and iii) the bias relative to observations ('bias vs. obs'). All values reported in this figure are computed over the set of validation stations.}}
    \label{tab:additional_models}
\end{table*}



\bibliographystyle{unsrtnat}
\bibliography{references}

\begin{figure*}[h!]
\centering
    \begin{subfigure}{\textwidth}
    \centering
    \includegraphics[width=0.83\textwidth]{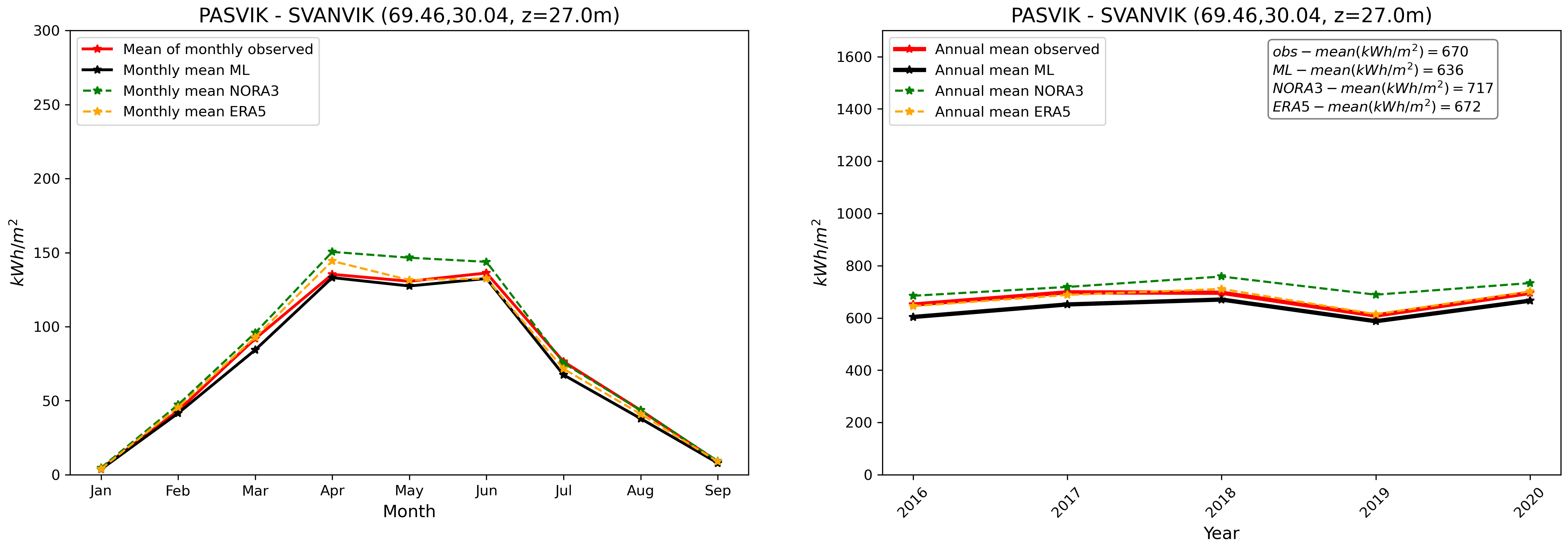}
    \includegraphics[width=0.83\textwidth]{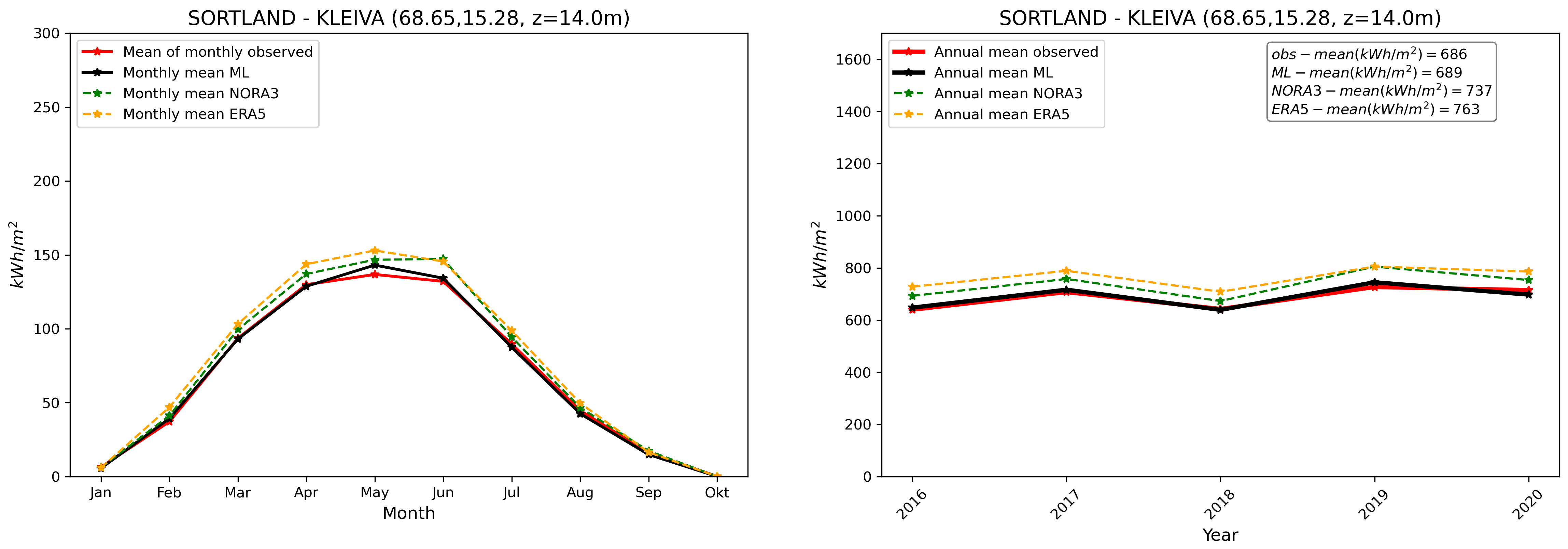}
    \includegraphics[width=0.83\textwidth]{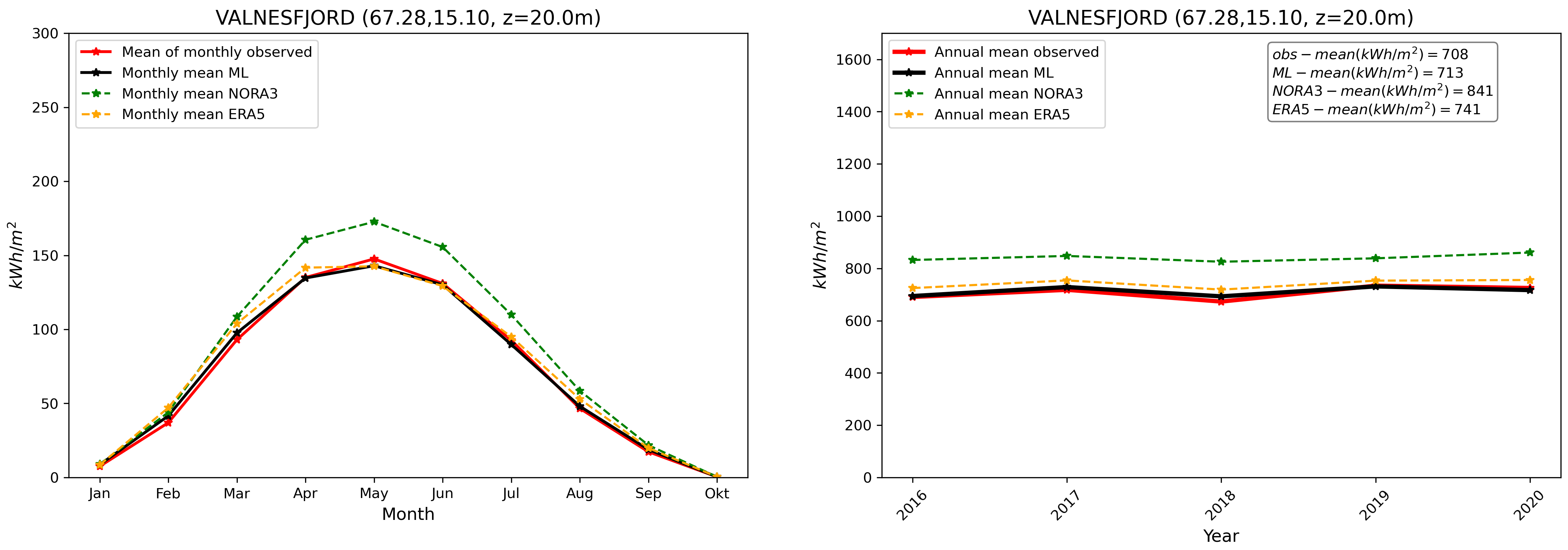}
    \includegraphics[width=0.83\textwidth]{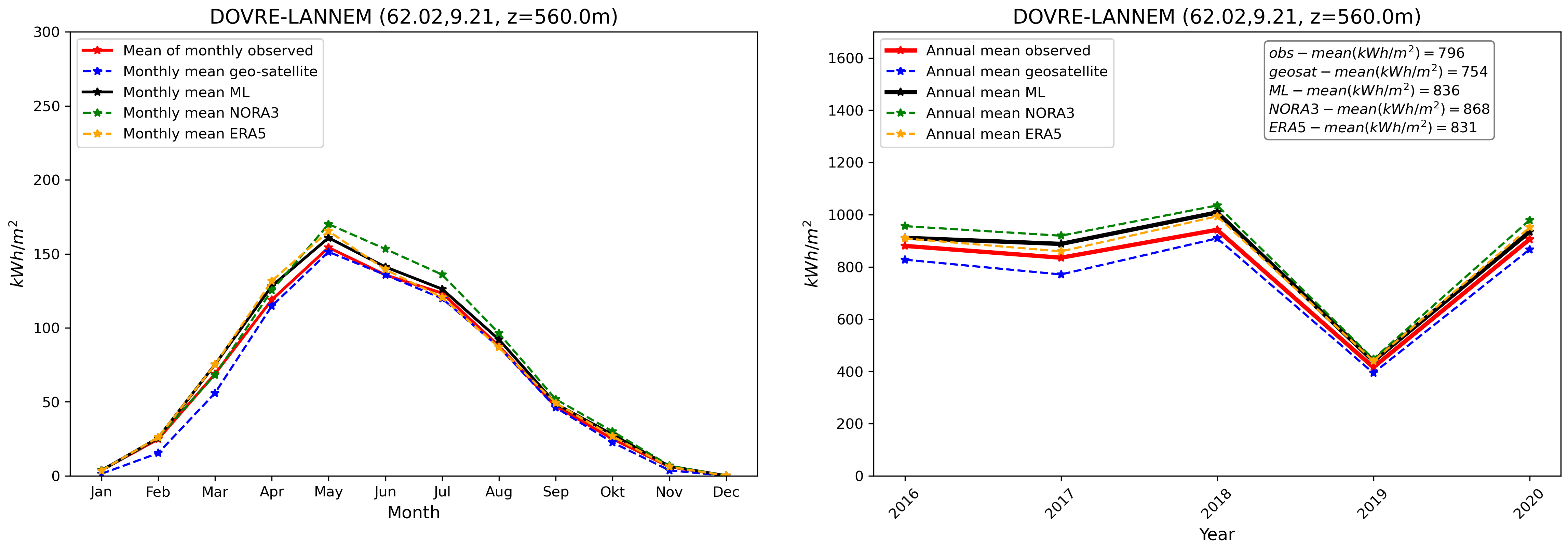}
    \end{subfigure}
    \caption*{(the figure continues on next page)}
\end{figure*}

\begin{figure*}[h!]
\centering
    \ContinuedFloat 
    \begin{subfigure}{\textwidth}
    \centering
    \includegraphics[width=0.83\textwidth]{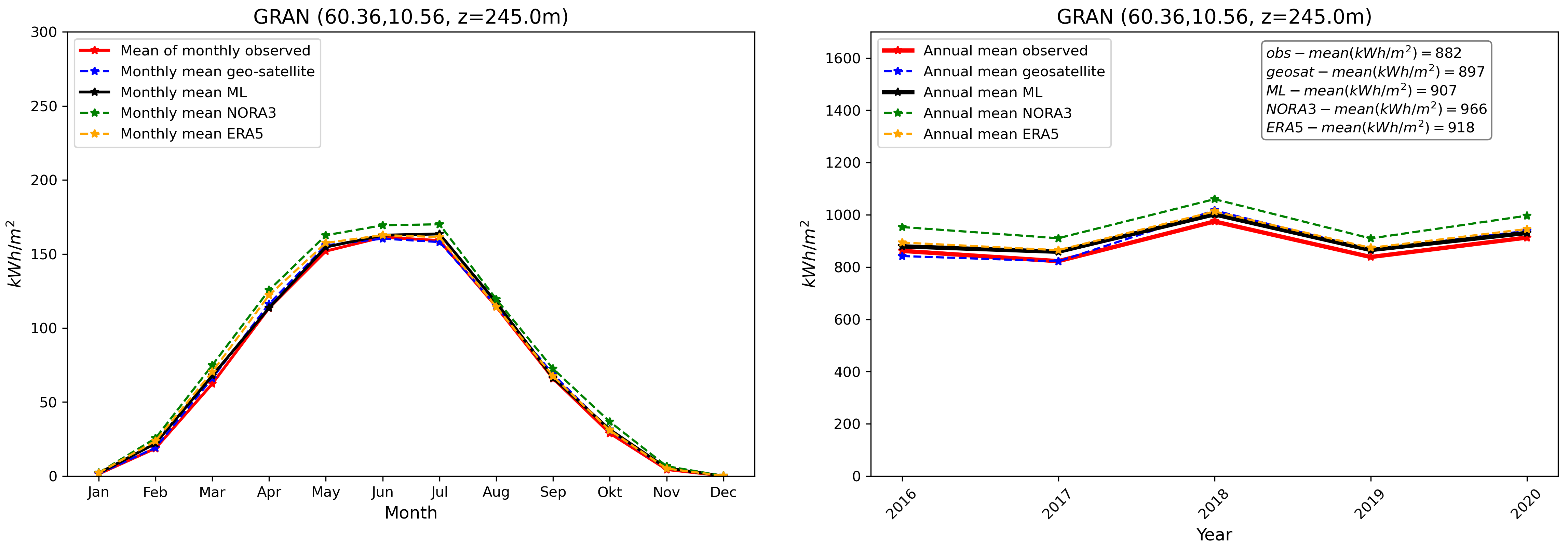}
    \includegraphics[width=0.83\textwidth]{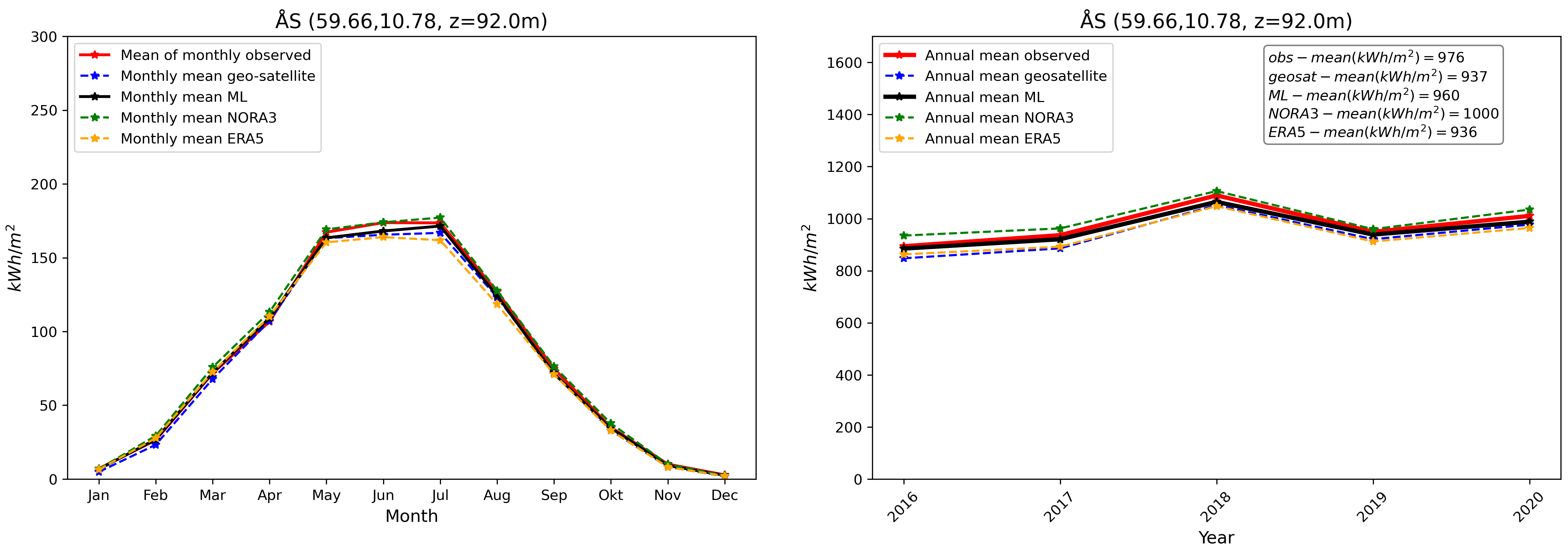}
    \includegraphics[width=0.83\textwidth]{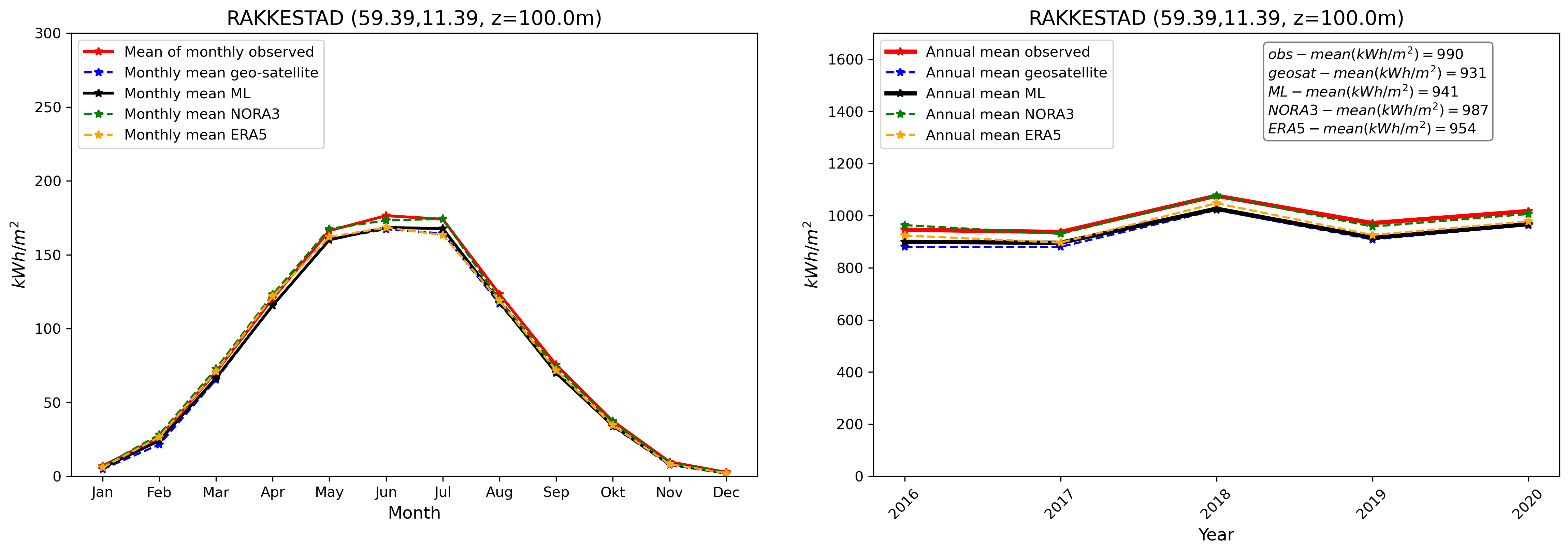}
    \includegraphics[width=0.83\textwidth]{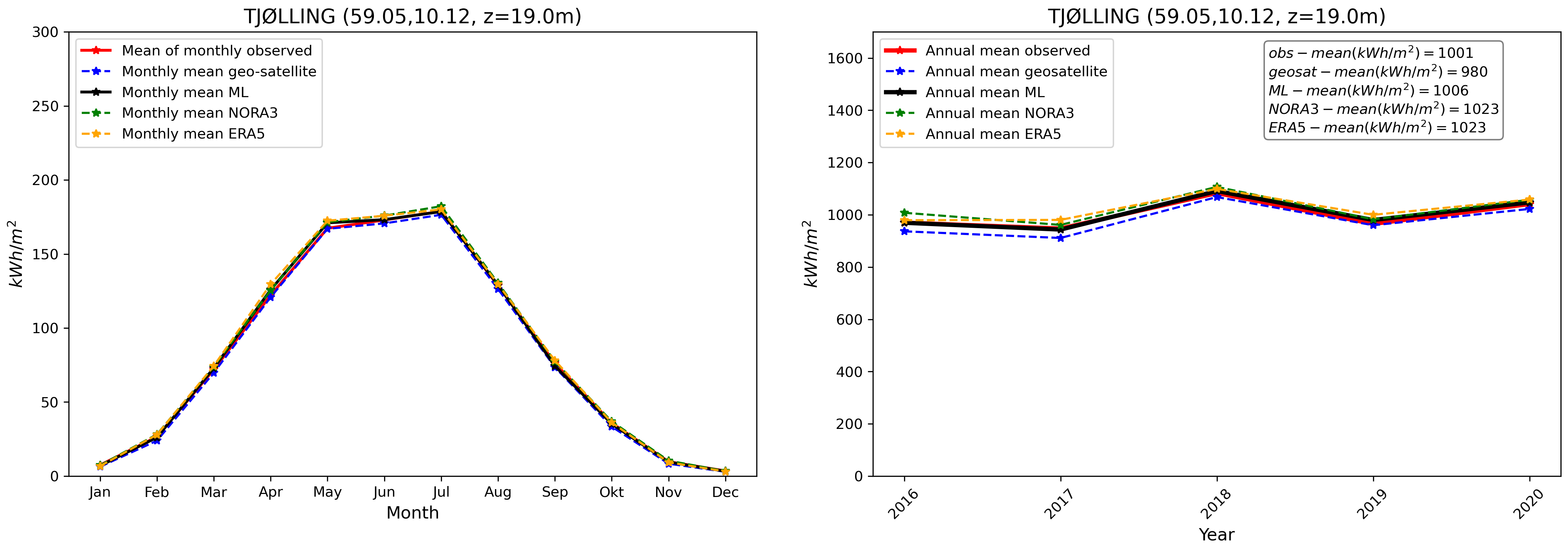}
    \end{subfigure}
    \caption{Comparison at the validation stations between the GHI estimates from our final ML product, other products (ERA5, NORA3, and SARAH-3 geosatellite data when these are available), and in-situ pyranometer observations. Our GHI product compares generally well with in-situ observations, in good agreement with the results presented in Fig. \ref{fig:taylor_plots}.}
    \label{fig:example_comparisons}
\end{figure*}

\end{document}